\RequirePackage{lineno}
\documentclass[12pt]{article}

\usepackage{times}
\usepackage{epsfig}
\usepackage{graphicx}
\usepackage{amsmath}
\usepackage{amssymb}
\usepackage{amsmath, amsthm}
\usepackage[margin=.75in]{geometry}
\usepackage[tiny,compact]{titlesec}
\usepackage{url}

\begin{document}
\renewcommand\linenumberfont{\normalfont\bfseries\small}


\title{Exploiting the Brain's Network Structure for Automatic Identification of ADHD Subjects}

\author{Soumyabrata Dey$^{1}$\thanks{Soumyabrata Dey, Computer Vision Lab., EECS Department, University of Central Florida, Orlando, Florida 32816-2362, USA, soumyabrata.dey@knights.ucf.edu}, A. Ravishankar Rao$^{2}$ and Mubarak Shah$^{1}$ \\
$^{1}$Computer Vision lab, EECS Department, University of Central Florida, Orlando, FL, USA; \\
$^{2}$IBM T.J. Watson Research Center, Yorktown Heights, NY, USA }
\maketitle

\vspace{-0.1in}
\begin{abstract}
Attention Deficit Hyperactive Disorder (ADHD) is a common behavioral problem affecting children. In this work, we investigate the automatic classification of ADHD subjects using the resting state Functional Magnetic Resonance Imaging (fMRI) sequences of the brain. We show that the brain can be modeled as a functional network, and certain properties of the networks differ in ADHD subjects from control subjects. We compute the pairwise correlation of brain voxels' activity over the time frame of the experimental protocol which helps to model the function of a brain as a network. Different network features are computed for each of the voxels constructing the network. The concatenation of the network features of all the voxels in a brain serves as the feature vector. Feature vectors from a set of subjects are then used to train a PCA-LDA (principal component analysis-linear discriminant analysis) based classifier. We hypothesized that ADHD-related differences lie in some specific regions of the brain and using features only from those regions is sufficient to discriminate ADHD and control subjects. We propose a method to create a brain mask that includes the useful regions only and demonstrate that using the feature from the masked regions improves classification accuracy on the test data set. We train our classifier with $776$ subjects and test on $171$ subjects provided by The Neuro Bureau for the ADHD-$200$ challenge. We demonstrate the utility of graph-motif features, specifically the maps that represent the frequency of participation of voxels in network cycles of length 3. The best classification performance (69.59\%) is achieved using 3-cycle map features with masking. Our proposed approach holds promise in being able to diagnose and understand the disorder.
\end{abstract}


\begin{figure*}[t]
\begin{center}
\begin{tabular}{c}
   \includegraphics[width=1\linewidth]
                   {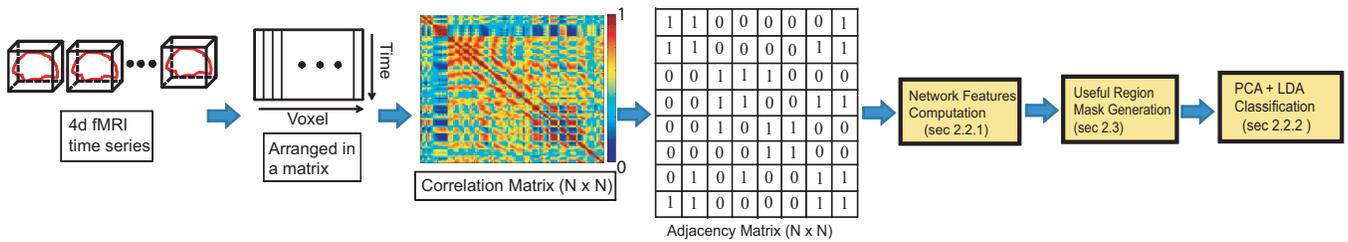}                   
\end{tabular}
\caption{Overview of our approach: Compute an $N \times N$ correlation matrix (N is the number of voxels ) using fMRI data; Compute the adjacency matrix by thresholding the low correlation values to generate a network; Compute network features such as node degree and cycle count for each node of the network; Generate a mask for the brain regions which are believed to be most effective for classification; Extract feature values within the generated brain mask and classify subjects using the PCA-LDA classifier.} \label{Fig_FlowChart}
\end{center}
\end{figure*}

\vspace{-0.1in}
\section{Introduction}
\label{sec:intro}

Attention Deficit Hyperactivity Disorder (ADHD) is a common behavioral disorder affecting children. Approximately 3-5\% of school aged children are diagnosed with ADHD. Currently, no well known biological measure exists to diagnose ADHD. Instead doctors rely on behavioral symptoms to identify it. To understand the cause of the disorder more fundamentally, researchers are using new structural and functional imaging tools like MRI and fMRI. fMRI has been widely used to study the functioning of brain. It provides high quality visualization of spatio-temporal activity within a brain, which can be used to compare the functioning of normal brains against those with disorders.

fMRI has been used for different functional studies of brain. Some of the researchers have used task-related fMRI data, in which the test subjects perform conscious tasks depending on the input stimuli. Others used resting state brain fMRI data. The brain remains active even during rest, when it is not engaged in an attentive task. Raichle et al. \cite{Raichle2001} identified several brain areas such as the MPFC, PCC and precuneus that are active during rest. These areas form part of a functional network known as the
resting-state network or default mode network (DMN) \cite{Damoiseaux2006}, \cite{Greicius2004}. The literature \cite{Cherkassky2006}, \cite{Damoiseaux2006}, \cite{Greicius2004}  tends to use interchangeably the concepts of resting state brain networks and the DMN as defined by Raichle in \cite{Raichle2001}. We compare the brain regions that we have found in the current ADHD data set with the components of the DMN described by Raichle in \cite{Raichle2001}.
It is believed that the DMN may be responsible for synchronizing all parts of the brain's activity; disruptions to this network may cause a number of complex brain disorders \cite{Raichle2010}. Researchers have studied neural substrates relevant to ADHD related behaviors, such as attention lapses, and identified the DMN as the key areas to better understand the problem \cite{Weissman2006}. In this study we use the resting state brain fMRI data and hypothesize that the differences between ADHD conditioned and control brains lie in the variation of functional connections of DMN.

Many studies have been performed to identify functional differences related to ADHD. Most of the approaches use group label analysis to deduce the statistical differences between ADHD conditioned and control groups. Structural MRI analysis suggests that there are abnormalities in ADHD brains, specifically in the frontal lobes, basal ganglia, parietal lobe, occipital lobe and cerebellum \cite{Castellanos1996,Overmeyer2001,Sowell2003,Seidman2006}. In another set of studies, ADHD brains were analyzed using task-related fMRI data. Bush et al. \cite{Bush1999} found significant low activity in the anterior cingulate cortex when ADHD subjects were asked to perform the Counting Stroop during fMRI. Durston et al. \cite{Durston2003} showed that ADHD conditioned children have difficulty performing the go/nogo task and display decreased activity in the frontostriatal regions. Teicher et al. \cite{Teicher2000} demonstrated that boys with ADHD have higher T2 relaxation time in the putamen which is directly connected to a child's capacity to sit still. A third set of work was done using the resting state brain fMRI to locate any abnormalities in the DMN. Castellanos et al. \cite{Castellanos2008} performed Generalized Linear Model based regression analysis on the whole brain with respect to three frontal foci of DMN, and found low negative correlated activity in precuneus/anterior cingulate cortex in ADHD subjects. Tian et al. \cite{Tian2006} found functional abnormalities in the dorsal anterior cingulate cortex; Cao et al. \cite{Cao2006} showed decreased regional homogeneity in the frontal-striatal-cerebellar circuits, but increased regional homogeneity in the occipital cortex among boys with ADHD. Zang et al. \cite{Zang2007} verified decreased Amplitude of Low-Frequency Fluctuation (ALFF) in the right inferior frontal cortex, left sensorimotor cortex, bilateral cerebellum, and the vermis, as well as increased ALFF in the right anterior cingulate cortex, left sensorimotor cortex, and bilateral brainstem.

While group level analysis can suggest statistical differences between two groups, it may not be that useful for clinical diagnosis at the individual level.  There have been relatively few investigations at the individual level of classification of ADHD subjects \cite{Dey2012, Dey2014, Dey2014-2, Dey9072496}. One such study is performed by Zhu et al. \cite{Zhu2010} who used a PCA-LDA based classifier to separate ADHD and control subjects at individual level. Unlike our network connectivity feature, which can connect all the synchronous regions of the whole brain, they used a regional homogeneity based feature for classification. Also the experiments were performed on only $20$ subjects, which are not conclusive.

Our algorithm exploits the topological differences between the functional networks of the ADHD and controlled brains. The different steps of our approach are described in the Fig \ref{Fig_FlowChart}. The input to our algorithm is brain fMRI sequences of the subjects. fMRI data can be viewed as a 4-D video such that the 3-D volume of the brain is divided into small voxels and imaged for a certain duration . The data can also be viewed as a time series of intensity values for each of the voxels. The correlation of these intensity time-series can be an indication of how synchronous the activities of two voxels are, and higher correlation values suggest that two voxels are working in synchronization. A functional network structure is generated for the brain of each of the subjects under study by computing the correlations for all possible pairs of voxels and establishing a connections between any pairs of voxels if their correlation value is sufficiently high. Different network features, such as degree maps, cycle maps and weight maps are computed from the network to capture topological differences between ADHD and control subjects. We have provided a detailed description of all the network features in the later sections of the article. A brain mask is computed that includes only the regions with useful information to classify ADHD and control subjects. For the rest of the article, we refer to this mask as a 'useful region mask'. The details of the useful region mask computation procedure are described in \ref{subsec:Method}. Finally, the network features from the voxels within the useful region mask are extracted to train a PCA-LDA based classifier. We have tested the performance of each of the network features computed on the training data set from the Kennedy Krieger Institute. We selected two different kinds of network features, degree map and 3-cycle map, for the experiments on the full data set.

In our work, we have performed experiments on a large challenging data set which includes subjects from different races, age groups, and data capturing sites. We propose a new approach for the automatic classification of ADHD subjects, and believe that our work will be helpful to the medical imaging community.

\begin{table*}\label{data-summery}
\begin{center}\scriptsize{
\begin{tabular} {| c | c | c | c | c | c | c | c | c |}   \hline

\textbf{Center} & \textbf{Sub Cnt} & \textbf{Age (yrs.)} & \textbf{Male} & \textbf{Female} & \textbf{Control} & \textbf{Combined} & \textbf{Hyperactive} & \textbf{Inattentive}
\\ \hline \multicolumn{9}{|c|}{\textbf{Training Data Set}}
\\ \hline \text{Kennedy Krieger Institute} & \text{$83$} & \text{$8$-$13$} & \text{$46$} & \text{$37$} & \text{$61$} & \text{$16$} & \text{$1$} & \text{$5$}
\\ \text{Neuro Image Sample} & \text{$48$} & \text{$11$-$22$} & \text{$31$} & \text{$17$} & \text{$23$} & \text{$18$} & \text{$6$} & \text{$1$}
\\ \text{New York University} & \text{$222$} & \text{$7$-$18$} & \text{$145$} & \text{$77$} & \text{$99$} & \text{$77$} & \text{$2$} & \text{$44$}
\\ \text{Oregon Health \& Sci. Univ.} & \text{$79$} & \text{$7$-$12$} & \text{$43$} & \text{$36$} & \text{$42$} & \text{$23$} & \text{$2$} & \text{$12$}
\\ \text{Peking University} & \text{$152$} & \text{$8$-$17$} & \text{$102$} & \text{$50$} & \text{$93$} & \text{$22$} & \text{$0$} & \text{$37$}
\\ \text{University of Pittsburg} & \text{$89$} & \text{$10$-$20$} & \text{$46$} & \text{$43$} & \text{$89$} & \text{$0$} & \text{$0$} & \text{$0$}
\\ \text{Wash. Uni. in St. Louis} & \text{$61$} & \text{$7$-$22$} & \text{$33$} & \text{$28$} & \text{$61$} & \text{$0$} & \text{$0$} & \text{$0$}
\\ \hline \multicolumn{9}{|c|}{\textbf{Test Data Set}}
\\ \hline \text{Kennedy Krieger Institute} & \text{$11$} & \text{$8$-$12$} & \text{$10$} & \text{$1$} & \text{8} & \text{3} & \text{0} & \text{0}
\\ \text{Neuro Image Sample} & \text{$25$} & \text{$13$-$26$} & \text{$12$} & \text{$13$} & \text{14} & \text{11} & \text{0} & \text{0}
\\ \text{New York University} & \text{$41$} & \text{$7$-$17$} & \text{$28$} & \text{$13$} & \text{12} & \text{22} & \text{0} & \text{7}
\\ \text{Oregon Health \& Sci. Univ.} & \text{$34$} & \text{$7$-$12$} & \text{$17$} & \text{$17$} & \text{27} & \text{5} & \text{1} & \text{1}
\\ \text{Peking University} & \text{$51$} & \text{$8$-$15$} & \text{$32$} & \text{$19$} & \text{27} & \text{9} & \text{1} & \text{14}
\\ \text{University of Pittsburg} & \text{$9$} & \text{$14$-$17$} & \text{$7$} & \text{$2$} & \text{5} & \text{0} & \text{0} & \text{4}
\\ \text{Brown University} & \text{$26$} & \text{$8$-$18$} & \text{$9$} & \text{$17$} & \text{-} & \text{-} & \text{-} & \text{-}
\\ \hline

\end{tabular}}
\caption{Summary of the data set released for ADHD-200 competition. A total of eight centers contributed to the data. The labels of the Brown University test set are not yet released.}
\end{center}
\end{table*}

\section{Materials and Method}\label{sec:Materials and Method}

\subsection{Data}\label{subsec:Data}
We use the data provided by the Neuro Bureau for the ADHD 200 competition which consists of $776$ training subjects and $197$ test subjects. Eight different centers contributed to the compilation of the whole data set, which makes the data diverse as well as complex. Different phenotypic information, such as age, gender, handedness, IQ, is also provided for each subject. Consider Table \ref{data-summery} for an overview of the data set. All research conducted by ADHD-200 data contributing sites was conducted with local IRB approval, and contributed in compliance with local IRB protocols. In compliance with HIPAA Privacy Rules, all data used for the experiments of this article is fully anonymized. The competition organizers made sure that the 18 patient identifiers are removed, as well as face information.

For all our experiments we have used preprocessed resting state fMRI data registered in a $4 \times 4 \times 4$ mm voxel resolution Montreal Neurological Institute (MNI) space, with nuisance variance removed, filtered using a bandpass filter (0.009 Hz \textless f \textless 0.08 Hz) and blurred with a $6$-mm FWHM Gaussian filter. All the fMRI scans are motion corrected to the first image of the time series. We have used a binary mask, provided with each of the subjects, to find out the voxels inside the brain volume. All the fMRI data volumes are of size $49 \times 58 \times 47$ voxels, but the number of sample across the time vary based on the center where data is captured. Further information regarding the data and the preprocessing steps is provided in \cite{ADHD200data}.

Though no quality control is performed on the data, a quality score is provided with each image file of all the subjects. The voxel-wise z-scores are thresholded and summed over all the voxels to compute the quality score of a image file. Images with low scores are considered to be better. We have not considered the quality scores for our study.

\vspace{0.1in}
\subsection{Method}\label{subsec:Method}
Network motifs such as node degree distribution, cycle etc. are analyzed in different disciplines of science to understand the systems being studied and neuroscience is not an exception \cite{Sporns2002}, \cite{Milo2002}, \cite{Ma'ayan2008}. We used different graph theoretic concepts for our study. We assume that the activity of a brain can be modeled as a functional network where the voxels are considered as the nodes, which are connected with each other based on the similarity of their activity over the time domain. In this article we have used the terms voxel and node interchangeably for the same meaning. The time series of a node is represented as a bold face notation. As the first step of the algorithm, we extract the time series for all the voxels and reorganized it as a separate 2-d matrix for each of the subjects in the data set. This is illustrated in second step of Figure \ref{Fig_FlowChart}. Next, the correlation between all possible voxel pairs is computed.  If a subject contains $N$ number of voxels, a correlation matrix of size $N \times N$ is constructed, where the $i^{th}$ row of the matrix corresponds to the pairwise correlation values of the $i^{th}$ voxel with all other voxels within the anatomical mask of the subject.

For any two voxels, if the time series are $\textbf{u}$ and $\textbf{v}$ respectively, the correlation can be computed as,
\begin{equation} \label{equationCorrelation}
r=\frac{(T \displaystyle\sum_{i=1}^{T} u_i v_i) - (\displaystyle\sum_{i=1}^{T} u_i)(\displaystyle\sum_{i=1}^{T} v_i)}{\sqrt{[T\displaystyle\sum_{i=1}^{T} u_i^{2}-(\displaystyle\sum_{i=1}^{T} u_i)^{2}][T\displaystyle\sum_{i=1}^{T} v_i^{2}-(\displaystyle\sum_{i=1}^{T} v_i)^{2}]}},
\end{equation}
where $T$ is the length of the time series, $\textbf{u} = [u_1, u_2, ... , u_T]$, $\textbf{v} = [v_1, v_2, ... , v_T]$.

We normalize all the time series between $[-1,$ $1]$ before correlation computation. Next, we threshold all the values of the correlation matrix to get a binary map of zeros and ones. This binary map can be considered as the adjacency matrix of a graph where the $i^{th}$ voxel is connected to all the voxels for which non-zero values are present in the $i^{th}$ row of the matrix. Note that we can consider two voxels to be connected by an edge when the correlation is high positive, high negative or simply the absolute value of the correlation is high. We have computed three different networks considering high positive, high negative and high absolute correlation values respectively.

\begin{figure}[tbp]
\begin{center}
\begin{tabular}{c}
   \includegraphics[width=0.3\linewidth]
                   {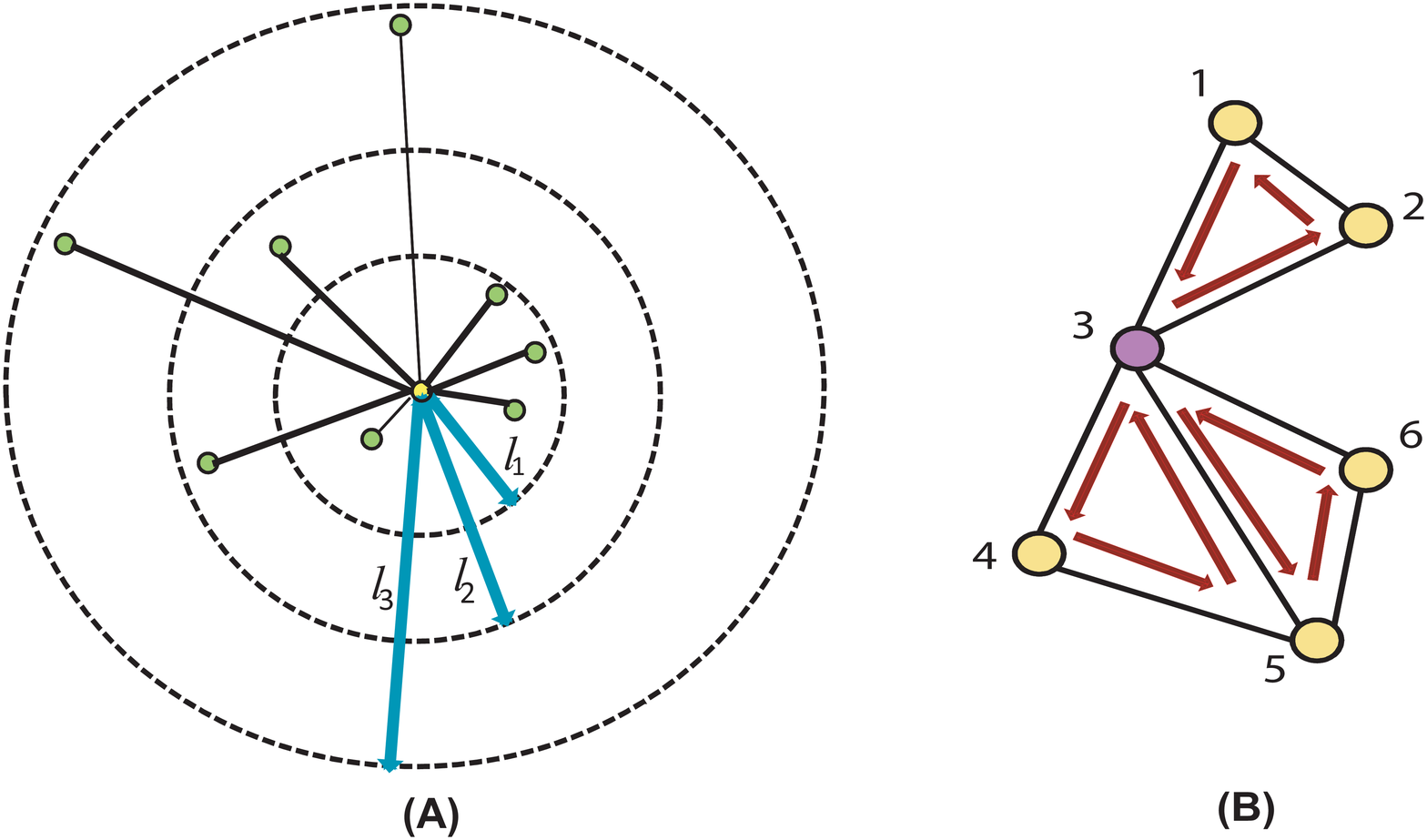}
\end{tabular}
\caption{\textbf{(A)} The degree of the node, highlighted in yellow, is the count of all the green nodes connected to it (i.e. $8$), while the varying distance degree is the counts of all the connected nodes in each of the bins defined by the three edge length thresholds ($l_1, l_2, l_3$) marked in blue.  In this example the varying distance degrees of the yellow node are $\{4, 2, 2\}$. \textbf{(B)} Shows all the distinct 3-cycles that containing the node $3$. } \label{Fig_networkFeatures}
\end{center}
\end{figure}

\vspace{0.1in}
\subsubsection{Network Feature Computation}\label{FeatureCompputation}
Once the graphs are constructed, for each subject of the data set, we compute different network features which can provide certain functional differences between the activity patterns of ADHD and control subjects' brain. The feature values from all the voxels of a network construct the feature map such as Degree Map, Cycle Map etc. The descriptions of different network features computed are given below.

\textbf{Degree:} For each node in a network, the degree is the count of the other nodes it is connected to. In other words, the degree of a node is the number of edges attached to it.

\textbf{Varying Distance Degree:}
Instead of considering the count of all the edges of a node as its degree, we group the edges based on their physical length and compute a separate degree for each of the groups. So, if we have $n$ threshold values for edge length, say $\{l_1, l_2, ..., l_n\}$, we can compute $n$ degrees, $\{d_1, d_2, ..., d_n\}$, of a node $v$, where $d_i$ is the count of all the edges connected to $v$ with length between $l_{i-1}$ to $l_i$. Refer to the Figure \ref{Fig_networkFeatures} for details. We use the Euclidian distance measure for the calculation of edge length. For the experiments, we have used threshold values 20, 40, and 80  and mm. where the average brain volume is approximately of size $172\times140\times140$ mm. Hence, we get 4 degrees per node which count edges of length 0-20 20-40, 40-80 and greater than 88 mm. respectively. The thresholds are selected through an intuitive basis such that different degrees should capture local to global connectivity pattern. The average percentage of degrees from close to far range are found as $70.44\%$, $16.54\%$, $8.40\%$ and $4.62\%$.

\textbf{L-cycle Count:}
A path in a network is a sequence of distinct nodes which can be traversed in the given order using the connecting edges. A cycle, on the other hand, is a closed path in the network where the starting and ending node is the same and all other nodes are distinct. The  L-cycle count of a node is the number of all possible distinct $L$ length cycles containing the node. Figure \ref{Fig_networkFeatures} illustrates this idea. L-cycle count for a node is calculated by traversing through all the L-length path starting from the node and counting the paths which leads to the starting node. The traversing can be performed using the breadth first search algorithm. We have used different cycle lengths for our experiments.

\textbf{Weight Sum:}
Instead of constructing an adjacency matrix using a threshold on the correlation values, we assume every node is connected to all other nodes by the weighted edges. The weight of the connecting edge of a node pair is their correlation value. As the correlation values can be positive and negative, we can separately add up all the positive, negative and absolute edge weights of a node to get its sum of positive, negative and absolute weights.

\vspace{0.1in}
\subsubsection{PCA-LDA Classification}\label{subsubsec:PCA-LDA Classification}
Once we finish computation of the network features, we extract the features from all of the voxels within the useful region mask. The mask generation algorithm is described in the next subsection. Concatenation of the feature values extracted from all the voxels generates a feature vector per subject. A PCA-LDA based classifier is trained separately using different set of feature vectors computed for different types of network features. Finally, the classifier is used for automatic classification of the ADHD subjects.

It is expected that the characteristics of the networks computed are represented by their feature vectors. A feature vector of a network represents a point in the feature space where the dimensionality of the space is same as the length of the vector. If the feature vectors of ADHD and control subjects are separable then their corresponding points in the feature space should cluster in different locations. When a classifier is trained, it learns to partition the feature space in such a way that the feature vectors from each of the groups are ideally clustered in separate segments. Given a feature vector of a test example, the classifier can identify which segments of the feature space it belongs to and classify the test subject accordingly. Linear Discriminant Analysis (LDA) is a widely used data classification technique which maximizes the ratio of between-class variance to the within-class variance to produce maximal separability. Mathematically, the objective is to maximize the following function :

\begin{equation} \label{equationLDA}
J(w) = \frac{w^T S_B w}{w^T S_W w}
\end{equation}
where $S_B$ and $S_W$ are between class and within class scatter matrix, and can be formulated as follows:

\begin{equation}
S_B = \displaystyle\sum_{i=1}^{n_A} ({x_i}^{(A)} - \mu^{(A)}) ({x_i}^{(A)} - \mu^{(A)})^T  + \displaystyle\sum_{i=1}^{n_C} ({x_i}^{(C)} - \mu^{(C)}) ({x_i}^{(C)} - \mu^{(C)})^T
\end{equation}

\begin{equation}
S_W = (\mu^{(A)} - \mu^{(C)}) (\mu^{(A)} - \mu^{(C)})^T,
\end{equation}
$n_A$ and $n_C$ are the number of subjects, $\mu^{(A)}$ and $\mu^{(C)}$ are the mean feature vectors, ${x_i}^{A}$ and ${x_i}^{C}$ are the $i$th feature vectors of the ADHD and control group respectively.

In many cases, the dimension of the feature space becomes so high that the proper partitioning of the space is difficult. For example, in our case, the dimension of the feature space is the number of voxels within the useful region mask which is several thousands. Again, most of the dimensions do not contain any significant data variance. Principal Component Analysis (PCA) is a procedure to find out a set of orthogonal directions, called principal components, along which the variance of the data is maximum. It then projects the data into the smaller dimensional subspace composed of the principal components. The classifier can work efficiently on the subspace which is significantly smaller in dimension than the original feature space. We use first 40  and first 100 principal components for the experiments on KKI and full data set respectively as they cover more than $98\%$ of data variance. We have included a plot of principal component vs. percent of data variance in the supplementary materials. Refer to \cite{Abdi2010} for details about PCA.

\begin{figure*}[t]
\begin{center}
\begin{tabular}{c}
   \includegraphics[width=0.8\linewidth]
                   {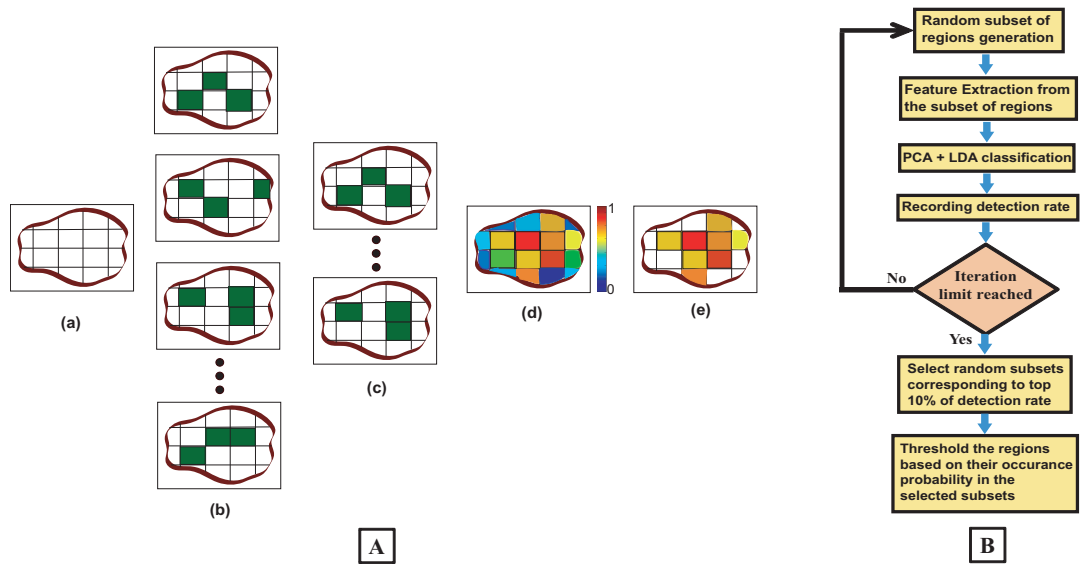}
\end{tabular}
\caption{ \textbf{(A)}This part of the figure explains the useful region mask generation algorithm on a single brain slice. The figure is just a graphical example, not the real data. In actual experiments brain volumes are used instead of slices and cube regions are used instead of square subdivision areas. \textbf{(a)} Divide the slice into square regions. \textbf{(b)} Select random sub sets of square regions marked in dark green. \textbf{(c)} Select the sub sets with top $10\%$ of detection rate. \textbf{(d)} Generate a probability map based on the regions occurrence in top $10\%$ subset. \textbf{(e)} Threshold the probability map to produce the useful region mask. \textbf{(B)} This part shows the flowchart for the mask generation algorithm.} \label{Fig_BrainMap}
\end{center}
\end{figure*}

\begin{figure*}[t]
\begin{center}
\begin{tabular}{c}
   \includegraphics[width=0.75\linewidth]
                   {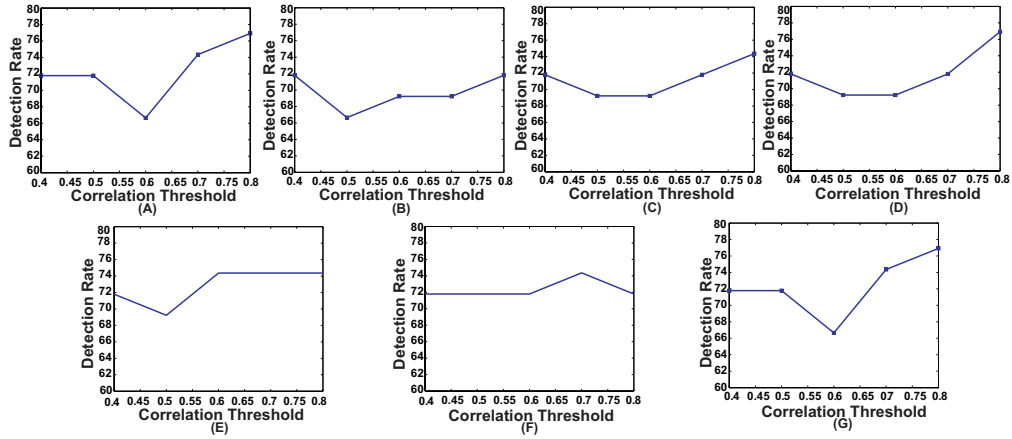}
\end{tabular}
\caption{The plots shows how detection rates for different network features change with correlation threshold. \textbf{(A)} Degree map positive correlations, \textbf{(B)} degree map negative correlations, \textbf{(C)} degree map absolute correlations, \textbf{(D)} varying distance degree map positive correlation, \textbf{(E)} $3$ cycle map positive correlation, \textbf{(F)} $4$ cycle map positive correlation, \textbf{(G)} weight map positive correlation.} \label{Fig_ROC_plots}
\end{center}
\end{figure*}

\begin{figure*}[t]
\begin{center}
\begin{tabular}{c}
   \includegraphics[width=0.7\linewidth]
                   {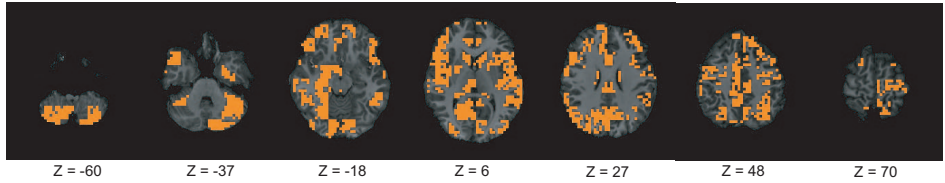}
\end{tabular}
\caption{The figure shows different slices to demonstrate the useful region mask computed. The masked regions are highlighted in orange color and overlaid on the structural images of a sample subject.} \label{Fig_usefulBrnMap}
\end{center}
\end{figure*}

\begin{table*}
\begin{center}\scriptsize{
\begin{tabular} {| c | c | c | c | c | c |}   \hline

\textbf{ROIs} & \textbf{[x, y, z] centers in \boldmath${mm.}$} & \textbf{ size in \boldmath${mm.^3}$ } & \multicolumn{3}{|c|}{\textbf{standard deviation in \boldmath${mm.}$}}
\\ \hline & & & \textbf{ x } & \textbf{ y } & \textbf{ z }
\\ \hline \text{Precuneus Cortex} & \text{[0, -66, 42]} & \text{7872} & \text{5.4894} & \text{6.6435} & \text{10.3592}
\\ \hline \text{Cingulate Gyrus} & \text{[0, -36, 52]; [0, 6, 42]} & \text{13056} & \text{4.5593} & \text{11.3751} & \text{10.9128}
\\ \hline \text{Temporal Pole} & \text{[56, 14, -18]} & \text{5312} & \text{4.7728} & \text{5.5878} & \text{5.7664}
\\ \hline \text{Superior Temporal Gyrus} & \text{[60, -18, -8]; [-60, -20, -4]} & \text{3392; 6400} & \text{7.1938; 6.6817} & \text{9.4413; 11.6393} & \text{4.0790; 5.7075}
\\ \hline \text{Inferior Temporal Gyrus} & \text{[54, -30, -20]; [-60, -48, -10]} & \text{1856; 2816} & \text{7.6293; 5.4892} & \text{6.7262; 8.2390} & \text{8.2617; 5.3582}
\\ \hline \text{Pre-central Gyrus} & \text{[-6, -22, 62]} & \text{8000} & \text{16.7226} & \text{8.5099} & \text{5.2886}
\\ \hline \text{Lingual Gyrus} & \text{[6, -64, 4]} & \text{19072} & \text{12.5240} & \text{11.4946} & \text{5.8835}
\\ \hline \text{Right Amygdala} & \text{[24, -2, -18]} & \text{2176} & \text{9.6639} & \text{7.3186} & \text{7.1020}
\\ \hline

\end{tabular}}
\caption{Shows list of the clusters and their approximate centers, sizes and standard deviations found using the most useful region mask algorithm. The coordinates are calculated on the HarvardOxford-cort-maxprob-thr0-1mm standard atlas provided with the FSL $4.1$. We list the ROIs of Harvard-Oxford Cortical and Subcortical Structural Atlases for which more than $50\%$ of the volumes are selected in the useful region mask. Atlas tool of FSL view is used for this purpose } \label{ROI list}
\end{center}
\end{table*}

\vspace{0.1in}
\subsection{Useful Region Mask}\label{subsec:brainMap}
Different research studies have proposed several Regions Of Interests (ROI) for fMRI analysis. These different ROIs vary in size and number. In some studies they are identified based on the anatomical structure of the brain and in other studies they depend on the functional responsibility. Tzourio-Mazoyer et al. \cite{TzourioMazoyer2002} identified the ROIs based on similar functional responses in the brain. Craddock et al. \cite{Craddock2011} generated a homogenous functional connectivity map from resting state fMRI data. Smith et al. \cite{Smith20072009} identified several co-varying functional subnetworks in the resting state brain. However, it is still unclear which ROIs are the best for resting state functional connectivity analysis. Also it is not known if all the ROIs detected by one method are required for ADHD classification or if the use of a subset of ROIs is more efficient. To find these answers, we use a novel method to identify the useful region mask for the classification of ADHD and control subjects.
The algorithm for the useful region mask generation is as follows:

\begin{description}
  \item[step 1] For each of the subjects, used for mask generation algorithm, we do the following:
                \begin{itemize}
                     \item Divide the brain volume into small cube-shaped regions. Each of the regions is typically $5 \times 5 \times 5$ voxels except the regions at the boundary of the brain volume.
                     \item Select a random subset of the regions. We include each region in the subset with probability \emph{p}.
                     \item Generate degree map by extracting the degrees for the voxels within the selected subset of regions.
                \end{itemize}
  \item[step 2] Train the PCA-LDA based classifier and calculate the detection accuracy on the test data set.
  \item[step 3] Perform the step 1 and step 2 for \emph{m} number of times, each time generating a different random subset, calculating the detection accuracy and recording it.
  \item[step 4] Choose the random sub sets corresponding to the top $10\%$ of the detection accuracy as the candidates for generating the useful region mask. We count the occurrence of each of the regions in all of the candidate sub sets and normalize the counts between 0 to 1 by dividing it by the number of candidate sub sets. This gives us the probability of inclusion of each of the regions in the mask.
  \item[step 5] Finally the useful region mask is generated using a threshold \emph{th} to prune the regions with low probability.
\end{description}

We experimentally verified that highest detection rate achieved when \emph{p} is 0.40 and \emph{th} is 0.60. The experiment results are included in the supplementary materials. The value of \emph{m} was kept as 500 so that the number of iterations should be large enough but computationally feasible. Figure \ref{Fig_BrainMap} (A) is an illustration of the proposed algorithm on a cartoon $2$-D slice of a brain while Figure \ref{Fig_BrainMap} (B) is the flowchart for the mask generation algorithm. Note that other network features may also be used in the algorithm but we simply use degree map feature. We assume that the regions, which are useful for identifying ADHD conditioned brains, should not vary depending on the feature used for the detection of the mask. We have tested the idea computing useful region mask using 3-cycle map feature also. We found that the final detection rates are very similar (check the supplementary materials).

\vspace{0.1in}
\section{Experiments and Results}\label{sec:Experiment and Results}
First, we verified the performance of each of the network features computed on a subset of the training data. We used fMRI data of $83$ subjects from the Kennedy Krieger Institute data set. Among the $83$ subjects, the first $44$ subjects are used for training and the remaining $39$ for testing. The performances of each of the features is computed with or without using the useful region mask. The mask is generated on the KKI training set comprising the first 44 subjects of the KKI subset and using the algorithm described in \ref{subsec:brainMap}. Each time a random subset of regions is selected, the classification performance is measured by leave-one-out cross verification, i.e. take $43$ subjects for training and test on the remaining one subject, repeat the process $44$ times, testing each of the $44$ subjects one at a time and averaging the correct detection count. Figure \ref{Fig_usefulBrnMap} shows the computed mask on different slices of the brain. Table \ref{ROI list} list the information of the different clusters found in the useful region mask and the ROIs they are overlapped with. To empirically select the correlation threshold to be used for our experiments, we varied it from $0.4$ to $0.8$ with an increment of $0.1$ in every step. In each step, detection rates for different network features are computed on the KKI test set of 39 subjects. The plots for correlation threshold vs. detection rate are shown in Figure \ref{Fig_ROC_plots}. To generate the plot for the weight map, we compute the sum of the edge weights considering only the edges which have weights greater than the correlation thresholds used within that step. Note that the detection rate for each feature is measured for positive, negative and absolute correlation values. However, the features computed from the positive correlation values have always outperformed the other two cases. Hence, we have not reported the other two cases in the paper. Since for all the network features, other than the 4-cycle map, the best performance is consistently achieved when correlation threshold is 0.80, we choose to use this value for all the experiments on the full data set.

\begin{table*}
\begin{center}\scriptsize{
\begin{tabular} {| c | c | c | c |}   \hline

\textbf{Feature} & \textbf{Correlation Threshold} & \textbf{Performance (\%)} & \textbf{Performance (\%)} \\
& & \textbf{using useful region mask} & \textbf{without useful region mask}

\\ \hline \text{Degree Map positive} & \text{$0.80$} & \text{$76.92$} & \text{$69.23$}
\\ \hline \text{Degree Map negative} & \text{$0.80$} & \text{$71.79$} & \text{$69.23$}
\\ \hline \text{Degree Map absolute} & \text{$0.80$} & \text{$74.36$} & \text{$71.79$}
\\ \hline \text{Varying Distance Degree Map} & \text{$0.80$} &  \text{$76.92$} & \text{$69.23$}
\\ \hline \text{$3$-cycle-map} & \text{$0.80$} & \text{$74.36$} & \text{$71.79$}
\\ \hline \text{$4$-cycle-map} & \text{$0.70$} & \text{$74.36$} & \text{$69.23$}
\\ \hline \text{Weight Map positive} & \text{$0.80$} & \text{$76.92$} & \text{$69.23$}
\\ \hline \text{BOW time series histogram} & \text{-} & \text{$69.23$} & \text{$66.67$}
\\ \hline \text{BOW Degree Map histogram} & \text{$0.80$} & \text{$69.23$} & \text{$66.67$}
\\ \hline \text{BOW time series and Degree Map histogram} & \text{$0.80$} & \text{$69.23$} & \text{$66.67$}

\\ \hline

\end{tabular}}
\caption{Initial test results shows the performance of all the network features computed on the Kennedy Krieger Institute's data set. Positive, negative and absolute keywords are used to indicate that positive, negative and absolute correlation values are considered for network generation. If any keyword is not specifically mentioned, then the positive correlation values are used only. } \label{initial-results}
\end{center}
\end{table*}

\begin{table*}
\begin{center}\scriptsize{
\begin{tabular} {| c | c | c | c | c | c | c | c | c | c | c | c | c |}   \hline

          & \multicolumn{3}{|c|}{\textbf{Deg. Map (mask)}}  & \multicolumn{3}{|c|}{\textbf{Deg. Map (no mask)}} & \multicolumn{3}{|c|}{\textbf{3-cycle Map (mask)}} & \multicolumn{3}{|c|}{\textbf{3-cycle Map (no mask)}}
\\ \hline  & \textbf{Det. Rate} & \textbf{Spec.} & \textbf{Sens.} & \textbf{Det. Rate} & \textbf{Spec.} & \textbf{Sens.} & \textbf{Det. Rate} & \textbf{Spec.} & \textbf{Sens.} & \textbf{Det. Rate} & \textbf{Spec.} & \textbf{Sens.}

\\ \hline \textbf{KKI} & \text{72.72} & \text{1} & \text{0} & \text{72.72} & \text{1} & \text{0} & \text{72.72} & \text{1} & \text{0} & \text{72.72} & \text{1} & \text{0}

\\ \hline \textbf{Neuro Image} & \text{68} & \text{.7857} & \text{.5454} & \text{64} & \text{.7143} & \text{.5454} & \text{72} & \text{.7857} & \text{.6364} & \text{68} & \text{.8572} & \text{.4545}

\\ \hline \textbf{NYU} & \text{70.73} & \text{.9167} & \text{.6207} & \text{65.85} & \text{.7500} & \text{.6207} & \text{70.73} & \text{.8333} & \text{.6552} & \text{63.41} & \text{.8333} & \text{.5517}

\\ \hline \textbf{OHSU} & \text{70.59} & \text{.7778} & \text{.4286} & \text{64.70} & \text{.7037} & \text{.4286} & \text{73.52} & \text{.8148} & \text{.4286} & \text{70.59} & \text{.7407} & \text{.5714}

\\ \hline \textbf{Peking} & \text{64.71} & \text{.8889} & \text{.3750} & \text{60.78} & \text{.8889} & \text{.2917} & \text{62.74} & \text{.9259} & \text{.2917} & \text{56.86} & \text{.9630} & \text{.1250}

\\ \hline \textbf{Pittsburgh} & \text{77.78} & \text{1} & \text{.5000} & \text{66.67} & \text{.8000} & \text{.5000} & \text{77.78} & \text{1} & \text{.5000} & \text{66.67} & \text{1} & \text{.2500}

\\ \hline \textbf{Overall} & \text{69.05} & \text{.8602} & \text{.4872} & \text{64.32} & \text{.7957} & \text{.4615} & \text{69.59} & \text{.8710} & \text{.4872} & \text{64.33} & \text{.8710} & \text{.3718}

\\ \hline
\end{tabular}}
\caption{Shows the detection rate, specificity and sensitivity of the classification experiments on the test data set released for the ADHD-200 competition. Comparison of the performances are shown when useful region mask is used and not used for the degree map and 3-cycle map features.} \label{fullDataSetResults}
\end{center}
\end{table*}

Table \ref{initial-results} summarizes the best performance obtained for each of the network features and the corresponding correlation threshold values. The performance in the table signifies the percentage of total number of correct detection (control and ADHD) among total number of test subjects. Note that for all the features, the performance without using useful regions mask is lower compare to when we use the mask. This demonstrate the utility of the voxel selection through the generated mask. In one of the recent studies Solmaz et al. used Bag of Word features for automatic classification of the ADHD subjects \cite{Solmaz2012}. We used their method for the purpose of comparison of the performances with our method. For our experiments using the Bag of Words feature, each subject is represented by $75$ and $100$ bin histograms when we used raw time series and degree map features respectively. A third kind of experiment performed by representing each of the subjects as a concatenation of two types of histograms resulting in a $175$ bin histogram. The details of the Bag of Word method are provided in the supplementary materials.

We perform thorough experiments on the full data set using positive degree map and positive 3-cycle map features. We trained our classifier with the full training data, which has $776$ subjects from $7$ different centers, and test on the $171$ subjects from $6$ centers released for the ADHD-$200$ competition. Again, we compared the performance with and without using the useful region mask. We reused the same mask generated using first 44 subjects of KKI. It is worth mentioning that the mask selects 6916 voxels from which features are extracted. The correct detection rate, specificity and sensitivity for each of the test centers and for overall centers are reported in Table \ref{fullDataSetResults}.
Since the subject labels of the Brown University test set have not yet been released, we cannot compute the performance measures on that subset.

\begin{figure*}[t]
\begin{center}
\begin{tabular}{c}
   \includegraphics[width=0.75\linewidth]
                   {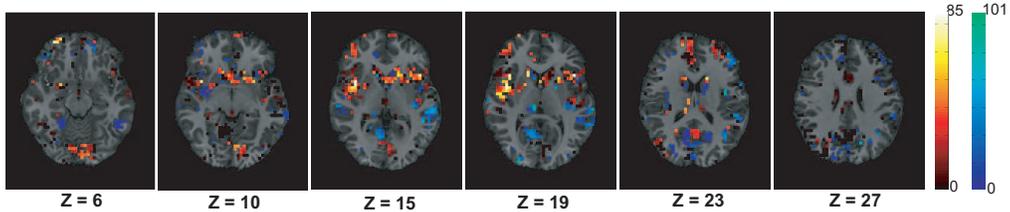}                   
\end{tabular}
\caption{The figure shows average difference of degrees of the control group from the ADHD group for the voxels within the useful region mask. The average difference  is calculated using the 83 subjects of KKI training set. The dark red to white color map is used to represents higher degree of control subjects and blue to green color map is used to show the opposite. The control group shows higher connectivity in the Cingulate Gyrus region on slices with Z coordinates $10$ and $15$ and Paracingulate Gyrus region on slices with Z coordinates $19$ and $23$.} \label{Fig_avgDegMap}
\end{center}
\end{figure*}

\vspace{0.1in}
\section{Discussion}\label{sec:Discussion}
We have modeled the brain as a functional network which is expected to represent the interaction of the different active regions of the brain. We assumed that ADHD is a problem caused due to the partial failure of the brain's communication network and the affected subjects can be distinguished from control subjects using the topological differences of their respective functional networks. To verify the idea, we have extracted different network features to train a PCA-LDA based automatic classifier. Figure \ref{Fig_avgDegMap} shows that the average degree map, computed for the ADHD and control subjects of the KKI data set, is able to capture some difference of connectivity in the Cingulate Gyrus and the Paracingulate Gyrus regions of brain. We also proposed that the features from the whole brain are not required for the classification, but some key areas hold useful information. Our results shows that the inclusion of features from the whole brain can negatively impact the classification accuracy. This resulted in a novel algorithm to compute the useful region mask which helped to improve the classification performance.

The different network features computed are expected to capture different characteristics of the functional network. The degree map and the weight map can capture how densely the nodes of the network is connected. This can give us a measure of how synchronously different regions of a brain are interacting. Varying distance degree map, on the other hand, can also reveal the fact that how the synchronous regions are distributed over the brain. While degree map only captures pairwise interactions of voxels, it ignores higher-order interactions, such as among three voxels simultaneously. We know from brain anatomy that there are such multiply connected brain regions. Hence, cycle maps offer a different perspective from which a given network may be viewed. The utility of using network motifs such as cycles to describe networks has been described in \cite{Milo2002}.

Figure \ref{Fig_usefulBrnMap} and Table \ref{ROI list} presents the ROIs found through our adaptive labeling technique described in Section \ref{subsec:brainMap}.  These ROIs were used in the classification including regions such as the cingulate and precuneus which is consistent with the findings of Castellanos et al. \cite{Castellanos2008}. The cingulate and precuneus regions are known to be part of the default-mode network \cite{Damoiseaux2006}. Many regions in the Table \ref{ROI list} have also been identified by Assaf et al. \cite{Assaf2010}, such as the precuneus, temporal pole, superior temporal gyrus, and pre-central gyrus. Regions in Table \ref{ROI list} that are consistent with those reported by Uddin et al. \cite{Uddin2009} include the inferior temporal gyrus and lingual gyrus. Interestingly, Table \ref{ROI list} identifies the right amygdala, which did not show up in the analysis of Castellanos et al. \cite{Castellanos2008} or Assaf et al. \cite{Assaf2010} or Uddin et al. \cite{Uddin2009}. The limbic system is known to play a role in ADHD, and a study by Plessen et al. \cite{Plessen2006} reported disrupted connectivity between the amygdala and OFC in the children with ADHD. Hence the value of our technique is that it provides an independent and automatic source of hypotheses about the brain regions that are implicated in the diagnosis and classification of ADHD. In this sense, our technique for ROI identification can be considered to be a model-free method. Furthermore, our classifier is agnostic to any particular theory of ADHD, and works strictly on a machine-learning approach to separating ADHD patients from controls by utilizing labeled data. Hence the technique described in this paper is applicable to other types of brain disorders where one can create labeled data for the accompanying brain scans.

The curves in Figure \ref{Fig_ROC_plots} show that for all the network features, high performance value is achieved when correlation threshold $0.80$ is used to construct the network. In four out of seven cases the performances are the highest, in other two cases they are one of the highest and in one case it is slightly lower that the highest. The results are not surprising since they indicate that the difference of connection structure for highly correlated voxels matters the most for classification.

Considering the results in Table 4, we observe that in 5 out of 6 data sets, the 3-cycle maps with voxel selection give the best detection rate.  Only on one data set, the Peking data set, the 3-cycle map with voxel selection gives marginally worse performance than the degree map with voxel selection. To the best of our knowledge, this is the first time that the utility of cycle-related features has been demonstrated in the fMRI imaging literature. The study in \cite{Ma'ayan2008} showed that cycle-related features are useful in discriminating biological networks from man-made networks, but did not investigate various types of fMRI-derived networks.

We note that calculating cycle-related features is more computationally intensive than the degree map, and the computation increases exponentially with cycle length. The use of GPUs can reduce the cost of computation, as earlier studies with fMRI images have shown \cite{Rao2011}. If standardized libraries for cycle computation become available on GPU platforms, it will promote the use of such features in fMRI research.

The use of the degree map provides a good compromise between classification performance and computational cost. It is easy to compute, and provides classification performance that is only marginally worse than that of the 3-cycle maps in most cases. One limitation of our study is that we have not used any specific measure to remove different signal to noise ratios which may be introduced in the data due to the difference of experimental setups among the sites. Also, some of the recent studies \cite{VanDijk2012} \cite{Power2012} indicate that the correlations of different brain regions are sensitive to the motion of the head even though the data is preprocessed for motion correction. We have not performed any explicit step to counter this problem. Finally, we note that we used a single classifier, the PCA-LDA method to investigate the utility of different network features. It is possible that other classifiers such as neural networks or support vector machines may give better performance. Such investigations need to be carried out in the future.

\subsection*{Acknowledgements}
The project described was supported by Award Number R21CA129263 from the National Cancer Institute.  The content is solely the responsibility of the authors and does not necessarily represent the official views of the National Caner Institute or the National Institutes of Health

Special thanks to The Neuro Bureau and all the data contributing sites for their efforts in compiling the large data set and making it publicly available. The goal of this project is that different disciplines of science may help to better understand the neural basis of ADHD.
\vspace{0.1in}

\bibliographystyle{IEEEtran}
\bibliography{report.bib}

\end{document}